\documentclass[sigconf,natbib=true]{acmart}
\usepackage[ruled,linesnumbered]{algorithm2e}
\usepackage{multirow}
\usepackage{makecell}
\usepackage{enumitem}
\usepackage{diagbox}
\setitemize[1]{noitemsep,partopsep=0pt,parsep=0pt,topsep=0pt, leftmargin=10pt,
rightmargin=0pt}
\AtBeginDocument{%
  \providecommand\BibTeX{{%
    \normalfont B\kern-0.5em{\scshape i\kern-0.25em b}\kern-0.8em\TeX}}}

\setcopyright{acmcopyright}
\copyrightyear{2018}
\acmYear{2018}
\acmDOI{10.1145/1122445.1122456}

\acmConference[Woodstock '18]{Woodstock '18: ACM Symposium on Neural
  Gaze Detection}{June 03--05, 2018}{Woodstock, NY}
\acmBooktitle{Woodstock '18: ACM Symposium on Neural Gaze Detection,
  June 03--05, 2018, Woodstock, NY}
\acmPrice{15.00}
\acmISBN{978-1-4503-XXXX-X/18/06}

\begin{document}

\title{Self-Balancing Gradient Allocation for Heterogeneity-Aware Feature Generation in Click-Through Rate Prediction}

\author{Moyu Zhang}
\affiliation{%
  \institution{Alibaba Group}
  \city{Beijing}
  \country{China}
}
\email{zhangmoyu@butp.cn}

\author{Yun Chen}
\affiliation{%
  \institution{Alibaba Group}
  \city{Beijing}
    \country{China}
}
\email{jinuo.cy@alibaba-inc.com}

\author{Yujun Jin}
\affiliation{%
  \institution{Alibaba Group}
  \city{Beijing}
  \country{China}
}
\email{jinyujun.jyj@alibaba-inc.com}

\author{Jinxin Hu}
\authornote{Corresponding Author}
\affiliation{%
  \institution{Alibaba Group}
  \city{Beijing}
  \country{China}
}
\email{jinxin.hjx@alibaba-inc.com}

\author{Yu Zhang}
\affiliation{%
  \institution{Alibaba Group}
  \city{Beijing}
  \country{China}
}
\email{daoji@alibaba-inc.com}

\author{Xiaoyi Zeng}
\affiliation{%
  \institution{Alibaba Group}
  \city{Beijing}
  \country{China}
}
\email{yuanhan@taobao.com}

\begin{abstract}
Generative pre-training via discrete diffusion provides dense reconstruction supervision across all feature fields simultaneously, yielding richer representations than discriminative CTR training alone. However, all existing generative CTR methods share a fundamental limitation: the reconstruction objective assigns equal training weight to every feature field, ignoring the profound heterogeneity of reconstruction difficulty across high-cardinality ID fields, sparse categorical attributes, numerical values, and behavioral sequences. This causes easy-to-reconstruct fields to dominate training gradients and leaves the hardest but most informative fields chronically underfit, a problem we identify as the \emph{generative difficulty imbalance}. Therefore, in this paper, we propose HeteGenCTR, a heterogeneous generative framework that resolves this imbalance through a single unified signal: per-field learnable difficulty parameters, jointly trained with the denoising network. This signal simultaneously drives two coordinated components with no additional hyperparameters. First, a self-balancing loss that automatically allocates more gradient budget to harder fields, with a provably stable equilibrium where each field's weight adapts inversely to its current reconstruction difficulty. Second, a difficulty-guided attention mechanism within the denoising network that suppresses the attention influence of already-converged easy fields and amplifies cross-field information flow toward hard fields. Both components are driven by the same learned difficulty signal, ensuring they remain mutually consistent throughout training. Extensive experiments demonstrate significant improvements over state-of-the-art generative baselines.
\end{abstract}



\maketitle

\section{Introduction}

Click-through rate (CTR) prediction is a fundamental task in industrial recommender systems, tasked with estimating the probability that a user will click on a presented item \cite{back1, back2, wdl, deepfm}. Standard CTR models are discriminative classifiers trained end-to-end with binary cross-entropy loss. While effective, this paradigm is inherently limited: it provides supervision only at the output scalar, leaving the model vulnerable to the severe data sparsity that characterizes real-world recommendation traffic \cite{emb1, rcola}. In production systems, the vast majority of user-item feature combinations appear only a handful of times, if at all, causing their learned representations to collapse or become unreliable during inference.

A compelling alternative paradigm frames CTR as a generative problem \cite{genctr, dgenctr, sgctr}. Rather than predicting a binary label directly, generative CTR models learn to reconstruct entire feature samples from a latent distribution, providing dense supervision across all feature fields simultaneously. This paradigm is typically implemented as a coupled two-stage workflow. In the first stage, a generative model is pre-trained to reconstruct feature fields. DGenCTR notably unifies the two stages by treating the click label as an additional feature field within the sample. The model is trained to reconstruct all fields including the label; when only the label is masked, the label reconstruction loss is mathematically equivalent to a CTR calibration loss, making CTR estimation a special instance of the generative denoising process. In the second stage, the pre-trained parameters and scoring function are directly inherited and fine-tuned for precise CTR prediction. This generative objective naturally regularizes feature representations: even rare feature values receive indirect supervision through their co-occurrence patterns with more frequent ones, mitigating the sparsity-driven collapse observed in purely discriminative training. Diffusion models, in particular, have demonstrated strong capacity for modeling complex high-dimensional distributions and have been applied to feature-level generation for CTR \cite{dgenctr, sgctr}, achieving promising gains over discriminative methods.

\begin{figure*}[t]
  \centering
  \includegraphics[width=\linewidth]{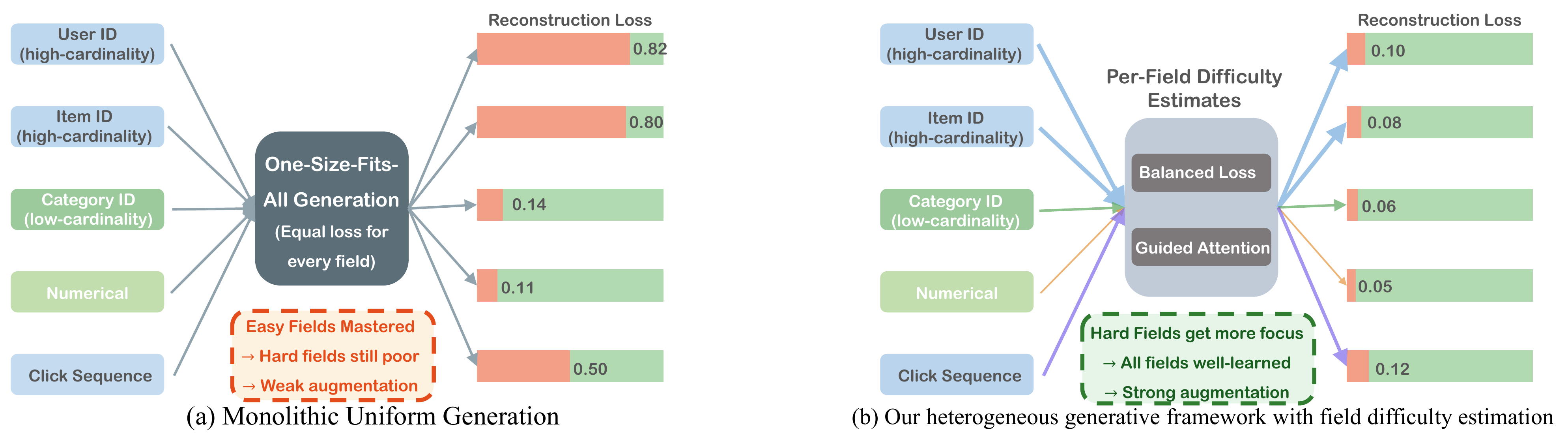}
  \caption{Illustration of the generative difficulty imbalance in CTR feature generation. (a) Monolithic uniform generation applies equal-weight loss across all feature fields, causing easy-to-reconstruct fields to dominate training gradients and suppressing learning for high-signal complex fields. (b) HeteGenCTR learns per-field difficulty estimates that automatically rebalance gradient budget and simultaneously drive difficulty-guided attention---all from a single unified signal.}
  \label{example}
  \vspace{-0.3cm}
\end{figure*}

Despite these advances, we identify a critical limitation shared by all existing generative CTR approaches: they treat feature generation as a \emph{homogeneous} task. This assumption is baked into the very formulation of their training objective---every feature field is assigned the same loss weight, regardless of its intrinsic reconstruction difficulty. In practice, this assumption is fundamentally violated by the structure of recommendation feature spaces.

CTR input features span radically different modalities and statistical properties: high-cardinality user and item ID fields with millions of unique values, sparse categorical attribute fields with moderate cardinality, dense numerical features, and variable-length sequential behavioral signals. These fields differ not only in their cardinality and sparsity but in their \emph{generative complexity}---the intrinsic difficulty of learning an accurate generative model for each field. A high-cardinality ID field requires the model to capture fine-grained identity patterns over a vast discrete space, while a numerical field may follow a compact distributional form. Categorical attributes with moderate cardinality present intermediate challenges, and sequence fields require temporal coherence across variable-length histories. Expecting a shared model to master all disparate modalities under a uniform training signal is unrealistic.

The consequences of this unrealistic assumption are severe. When all fields are trained with equal loss weight, the optimization dynamics inevitably become imbalanced. Fields that are easy to reconstruct---such as low-cardinality categorical attributes or near-constant numerical features---converge early and exert disproportionate gradient influence, pulling the shared model parameters toward representations that prioritize these simple fields. Meanwhile, high-complexity fields that are slow to converge---such as high-cardinality ID fields and behavioral sequences---remain underfit throughout training. We term this the \emph{generative difficulty imbalance}: the model attains good reconstruction quality on easy fields while failing to capture fine-grained patterns of the fields that actually carry the most predictive signal for downstream CTR task.

This is not merely a theoretical concern. ID fields and sequence fields are precisely the features that carry the strongest personalization signal in CTR prediction: user IDs encode long-term preference profiles, item IDs encode content identity, and click sequences encode short-term intent. When the generative model systematically underfits these high-signal fields because easy fields monopolize the training gradient, the resulting feature representations provide weak supervision for the most informative aspects of the feature space. The downstream CTR model therefore receives impoverished training signal exactly where it matters most.

Nor is this problem addressed by existing generative CTR methods. Both DGenCTR \cite{dgenctr} and SGCTR \cite{sgctr} follow a coupled two-stage paradigm, but DGenCTR notably dissolves the boundary between them: it treats the click label as an additional feature field to be reconstructed alongside the input features in the diffusion pre-training stage. When only the label is masked, the label reconstruction loss is mathematically equivalent to a CTR calibration loss, making CTR estimation a special instance of the generative denoising process. The second stage then directly inherits the pre-trained parameters and scoring function for supervised fine-tuning. Despite this architectural innovation, DGenCTR treats feature generation as a homogeneous task: it introduces per-field noise schedules in the diffusion process, but still sums per-field reconstruction losses with equal weights---leaving the gradient imbalance untouched. SGCTR applies a uniform generation objective across all fields. Prior multi-task learning techniques such as GradNorm and PCGrad \cite{gradnorm, pcgrad} are designed for discrete task boundaries with explicitly defined task outputs, and are not directly applicable to the continuous per-field difficulty variation encountered in feature generation.

We observe that the core problem is the absence of a \emph{per-field difficulty signal} that tells the model, for each field, how uncertain its reconstruction currently is. If such a signal were available, it could drive two natural corrections simultaneously: first, reweight the loss so that fields with high difficulty receive stronger gradient signal; second, modulate the attention mechanism in the denoising network so that easy fields do not drown out hard fields. Crucially, both corrections should derive from the same learned signal, without introducing any new hyperparameters to tune.

Therefore, in this paper, we propose \textbf{HeteGenCTR}, a heterogeneous generative framework with unified field difficulty estimation for CTR prediction, built on the discrete diffusion process. HeteGenCTR learns a single scalar difficulty estimate for each feature field, updated jointly with the denoising network during training. This unified signal drives two coordinated mechanisms. First, a self-balancing loss derived from multi-task uncertainty weighting: fields with high reconstruction difficulty retain large loss weights, while converged easy fields are progressively down-weighted. The equilibrium is provably stable, with a unique local minimum for each field's weight given the current reconstruction loss. Second, a difficulty-guided attention mechanism that scales each field's query in the HSTU denoising network by a difficulty-derived factor, suppressing easy-field attention and amplifying hard-field cross-field information flow without adding any new parameters. The attention scaling is derived to ensure that the effective attention weighting aligns with the loss-level weighting, so both mechanisms reinforce the same difficulty signal consistently. Both mechanisms share the same learned difficulty estimates and introduce no extra hyperparameters to tune.

The contributions of our paper are summarized as follows:
\begin{itemize}
\item We identify and formalize the \emph{generative difficulty imbalance} in generative CTR modeling, demonstrating that the uniform treatment of feature fields causes easy fields to dominate training gradients and suppress learning for high-signal ID and sequence fields. We further establish that existing per-field noise schedule adaptations in DGenCTR address a distributional mismatch orthogonal to this gradient imbalance, and that the two problems require independent solutions.
\item We propose HeteGenCTR, which introduces a unified per-field difficulty signal that simultaneously drives self-balancing loss aggregation and difficulty-guided attention modulation---two coordinated mechanisms with no additional hyperparameters beyond those in the baseline. We provide a stability analysis showing the self-balancing equilibrium is a strict local minimum, and a principled derivation showing the attention scaling aligns with the loss-level weighting by design.
\item Experiments across five datasets and an online A/B test confirm consistent, statistically significant improvements over state-of-the-art generative and discriminative CTR baselines, demonstrating that heterogeneity-aware generation is a critical and previously overlooked dimension of generative CTR modeling.
\end{itemize}

\section{Related Work}

\subsection{CTR Prediction Models}
Deep learning has driven substantial advances in CTR prediction through increasingly sophisticated feature interaction architectures. Early models such as WDL \cite{wdl} and DeepFM \cite{deepfm} jointly train a shallow memorization component---wide linear regression or factorization machines for sparse feature crosses---with a deep neural network for dense embedding generalization. DCN \cite{dcn} and its successor DCN-v2 \cite{dcn2} replace the wide component with explicit cross layers that learn bounded-degree feature interactions through residual-like compositions, with DCN-v2 further factorizing the cross weight matrices to improve parameter efficiency. Subsequent work pushes toward higher-order interactions: xDeepFM \cite{xdeepfm} introduces a compressed interaction network (CIN) that models vector-level $k$-th order crosses at the $k$-th layer; AutoInt \cite{autoint} casts feature interaction as multi-head self-attention over the feature embedding space; and MaskNet \cite{masknet} couples instance-guided feature selection with masked multi-head attention to adaptively highlight informative features per sample. FiBiNet \cite{fibinet} employs a Squeeze-and-Excitation network to reweight features before applying bilinear feature interactions, while GDCN \cite{gdcn} gates the outputs of cross layers and a deep tower to dynamically control their relative contributions. On the personalization front, PEPNet \cite{pepnet} modulates network parameters through domain-specific and user-specific embedding gates, and HSTU \cite{hstu} scales sequential transducers to trillion parameters under a generative recommendation formulation. Despite these architectural innovations, all these models operate within the discriminative paradigm, optimizing a binary classification objective end-to-end. This leaves them inherently vulnerable to data sparsity: feature combinations that appear rarely in training receive insufficient gradient signal, causing their representations to collapse or generalize poorly. The generative CTR paradigm was proposed precisely to address this limitation by providing field-level supervision through feature reconstruction.

\subsection{Generative Paradigms for CTR}
GenCTR \cite{genctr} introduces masked generative pre-training over feature tokens, demonstrating that generative supervision yields richer feature representations than purely discriminative training. DGenCTR \cite{dgenctr} applies discrete diffusion to model the joint distribution of categorical feature values and introduces per-field noise schedules, but retains a uniform sum over per-field losses---leaving the gradient imbalance problem unresolved. SGCTR \cite{sgctr} proposes a symmetric paradigm combining masked generative pre-training with discriminative fine-tuning. All these methods follow a coupled two-stage pipeline. DGenCTR in particular dissolves the boundary between the stages by representing each sample as $\boldsymbol{X}=\{\mathbf{F},y\}$ and training the diffusion model to reconstruct the click label alongside the input features. When only the label is masked, the label reconstruction loss is mathematically equivalent to a CTR calibration loss, making CTR estimation a special instance of the generative denoising process. The second stage then directly inherits the pre-trained scoring function for supervised fine-tuning. Nevertheless, all these methods treat feature generation as a homogeneous task, and none addresses the heterogeneity of generative difficulty across feature fields. Diffusion models \cite{ddpm, d3pm} have emerged as a powerful generative framework for recommendation, with applications in collaborative filtering \cite{diffrec} and sequential recommendation \cite{diff_aug}. In the CTR domain, discrete diffusion \cite{d3pm, mdm} is particularly well-suited due to the categorical nature of recommendation features. Our work builds on the discrete diffusion framework but extends it with a unified difficulty-driven mechanism that simultaneously rebalances gradient allocation at the loss level and modulates cross-field information flow at the attention level. HeteGenCTR directly addresses the shared limitation of these generative methods: the absence of a mechanism that recognizes and adapts to the heterogeneous reconstruction difficulty across feature fields.

\subsection{Multi-Objective Training Balance}
Training a shared network on multiple objectives of varying difficulty is a well-recognized challenge \cite{gradnorm, pcgrad, mgda, cagrad}. GradNorm \cite{gradnorm} dynamically adjusts gradient norms; PCGrad \cite{pcgrad} projects conflicting gradients; MGDA \cite{mgda} finds Pareto-optimal updates. These methods require per-task gradient access or Pareto optimization and are designed for discrete task boundaries.

A more principled alternative, proposed by Kendall~et~al.~\cite{uncertainty_mtl}, frames multi-task loss weighting as maximum likelihood estimation under homoscedastic (task-level) uncertainty. By modelling each task's output with a Gaussian or softmax likelihood parameterised by a task-specific noise scalar $\sigma_k$, they show that maximizing the joint log-likelihood over all tasks naturally yields a loss in which each task's loss is weighted by $1/\sigma_k^2$ and regularised by $\log \sigma_k$. This completely eliminates the need for hand-tuned loss weights. HeteGenCTR adapts and extends this probabilistic framework to the feature generation setting: we treat each feature field as an independent classification ``task'' with its own homoscedastic difficulty $\sigma^i$, derive the self-balancing loss from the joint maximum likelihood objective, and further extend the learned difficulty signal to simultaneously coordinate loss-level gradient rebalancing and attention-level information-flow modulation---creating a holistic, parameter-free solution to the multi-level difficulty imbalance in feature generation.

\section{Preliminary}

\subsection{CTR Prediction Task}
The CTR prediction task estimates the probability that a user will click on a presented item. It is formulated as a supervised binary classification problem. Given a feature set $\mathbf{F} = [f^1, f^2, \ldots, f^N]$ composed of $N$ feature fields and a binary label $y \in \{0, 1\}$, the task learns a function $\mathcal{F}: \mathbf{F} \rightarrow [0, 1]$ to estimate the click probability:
\begin{gather}
P(y | \mathbf{F}) = \mathcal{F}(f^1, f^2, \ldots, f^N)
\end{gather}
The feature fields encompass heterogeneous modalities: high-cardinality ID fields $\mathcal{F}^{ID}$, sparse categorical attribute fields $\mathcal{F}^{cat}$, dense numerical fields $\mathcal{F}^{num}$, and behavioral sequence fields $\mathcal{F}^{seq}$, i.e., $\{f^i\}_{i=1}^N = \mathcal{F}^{ID} \cup \mathcal{F}^{cat} \cup \mathcal{F}^{num} \cup \mathcal{F}^{seq}$.

\subsection{Generative CTR via Discrete Diffusion}

The generative CTR paradigm alleviates the limitation of discriminative binary classification by learning a generative model $p_\theta(\mathbf{F})$ of the joint feature distribution. Through field-level reconstruction pre-training, the model receives dense supervision and learns richer representations than binary classification alone. The pre-trained parameters and scoring function are directly inherited for CTR fine-tuning \cite{dgenctr, sgctr}.

\textbf{Discrete Diffusion and Reconstruction.} DGenCTR \cite{dgenctr} adopts an \emph{absorbing} discrete diffusion formulation \cite{d3pm}. For each field $i$, the forward process is a continuous-time Markov chain governed by a per-field masking rate $\gamma^i(t)$ toward an absorbing state $[\boldsymbol{M}]$. A denoising network $p_\theta$ learns the reverse process: given a partially masked feature set at timestep $t$, it predicts the original unmasked tokens. Numerical fields are discretized into $B$ uniformly spaced bins to apply the same formulation. Because high-cardinality ID features render full-vocabulary softmax intractable, DGenCTR approximates reconstruction via batch-softmax with cosine similarity over negatives drawn from the current batch \cite{dgenctr}:
\begin{gather}
q_{\theta}(\hat{e}^i \,|\, \boldsymbol{X}_t^{\setminus i}) = \frac{\exp(\cos(\hat{e}^i, G(\boldsymbol{X}_t^{\setminus i})))}{\sum_{\tilde{e}^i \in \mathcal{B}_i} \exp(\cos(\tilde{e}^i, G(\boldsymbol{X}_t^{\setminus i})))}
\label{eq:batchsoftmax}
\end{gather}
where $G(\cdot)$ is the scoring network and $\mathcal{B}_i$ is the batch-negative set.

\textbf{Coupled Two-Stage Pipeline.} Generative CTR frameworks follow a coupled two-stage workflow. DGenCTR dissolves the boundary between stages by treating the click label $y$ as the $(N+1)$-th feature field, so each sample is $\boldsymbol{X}=\{\mathbf{F},y\}$. The diffusion model is trained to reconstruct all fields including the label. When only the label is masked, the denoising network predicts $y$ from the unmasked features $\mathbf{x}_t^{\setminus y}$, yielding:
\begin{gather}
\mathcal{L}_{label} = -\log p_\theta(y | \mathbf{x}_t^{\setminus y})
\label{eq:label}
\end{gather}
parameterised by the same scoring function used for feature reconstruction:
\begin{gather}
p_\theta(y=1 | \mathbf{x}_t^{\setminus y}) = \frac{1}{1 + \exp\!\bigl(-(\mathcal{F}(y=1|\mathbf{F}) - \mathcal{F}(y=0|\mathbf{F}))\bigr)}.
\label{eq:score}
\end{gather}
Because the second-stage objective employs the exact same scoring function $\mathcal{F}(\cdot)$, CTR estimation is a special instance of the generative denoising process---the two stages are deeply coupled. The standard discrete diffusion training objective minimized in pre-training is:
\begin{gather}
\mathcal{L}_{gen} = -\mathbb{E}_t\!\left[\sum_{i=1}^N \log p_\theta(f^i_0 | \mathbf{x}_t, t)\right]
\label{eq:lgen}
\end{gather}
This formulation assigns equal weight to every feature field, regardless of cardinality, sparsity, or generative difficulty.

\begin{figure*}[t]
  \setlength{\abovecaptionskip}{-0.8cm}
  \centering
  \includegraphics[width=\linewidth]{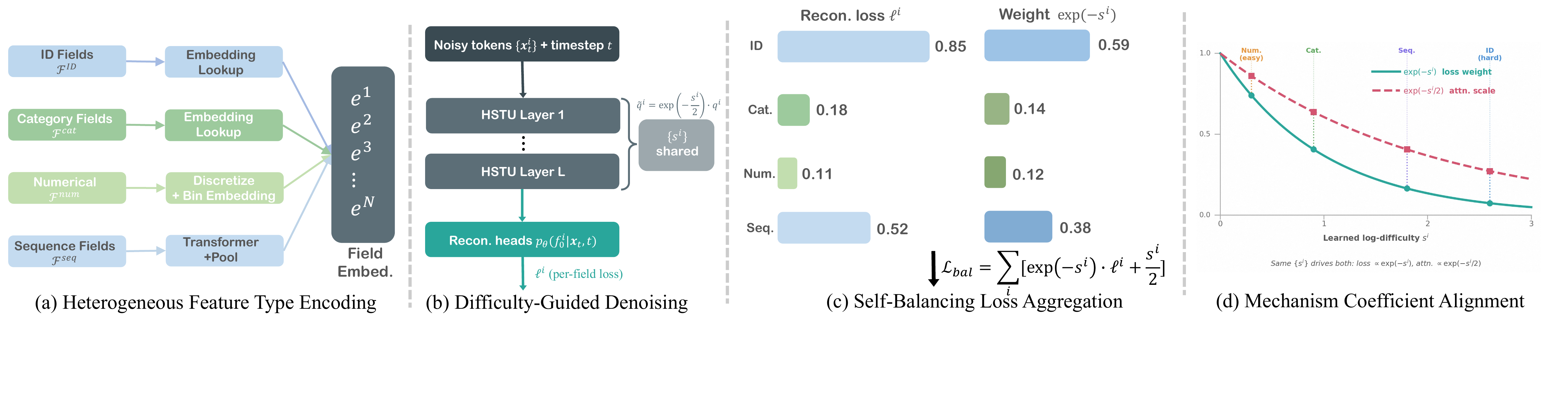}
  \caption{The overall architecture of HeteGenCTR. (a) Heterogeneous Feature Type Encoding partitions feature fields into four type groups. (b) The HSTU denoising network applies difficulty-guided attention scaling $\tilde{q}^i = \exp(-s^i/2)\cdot q^i$ using the same learned $\{s^i\}$. (c) Self-Balancing Loss aggregates per-field reconstruction losses weighted by $\exp(-s^i)$.}
  \label{model}
  \vspace{-0.5cm}
\end{figure*}

\textbf{Field Complexity and Noise Schedules.} DGenCTR introduces per-field noise schedules $\{\gamma^i(t)\}$ to accommodate varying vocabulary sizes. This engineering adaptation addresses distributional mismatch---larger vocabularies require slower masking rates---but does not affect the relative weighting in $\mathcal{L}_{gen}$; all fields still contribute with coefficient $1$. The gradient imbalance arises from loss aggregation, not the noise schedule. Because reconstruction operates at the embedding level, generative difficulty is determined by inferability: ID fields are hardest (unstructured memorization across millions of embeddings), sequence fields are intermediate (structured summary vectors), and categorical/numerical fields are easiest. HeteGenCTR retains per-field noise schedules while addressing the orthogonal problem of loss-level gradient imbalance.

\subsection{The Generative Difficulty Imbalance}\label{sec:motivation}

We characterize the generative difficulty of field $i$ by the convergence behavior of its reconstruction loss. Define the \emph{normalized reconstruction difficulty} $d^i(\tau)$ at training step $\tau$ as:
\begin{gather}
d^i(\tau) = \frac{\ell^i(\tau)}{\ell^i(0)}
\end{gather}
Empirically, high-cardinality ID fields (cardinality $> 10^5$) maintain $d^i(\tau) > 0.8$ throughout training under uniform weighting, while low-cardinality categorical attribute fields (cardinality $< 10^2$) converge to $d^i(\tau) < 0.2$ within the first $10\%$ of training. The gradient of the total loss $\mathcal{L}_{gen}$ is therefore dominated by easy fields during the bulk of training, suppressing the gradient signal for informative but hard-to-learn ID fields.

This imbalance manifests at multiple levels:
\begin{itemize}
\item \textbf{Loss level}: easy fields contribute disproportionately large gradients to $\nabla_\theta \mathcal{L}_{gen}$.
\item \textbf{Attention level}: within the denoising network, attended queries from converged easy fields tend to dominate the attention output, suppressing cross-field information flow toward hard fields.
\end{itemize}

A principled solution should address both levels simultaneously, ideally from a single, learned, per-field signal.

\section{Method}\label{sec:method}

We propose HeteGenCTR, a heterogeneous generative framework with unified field difficulty estimation for CTR prediction, built on discrete diffusion. HeteGenCTR operates entirely within the feature-reconstruction portion of the first stage: it improves the quality of feature reconstruction, and the resulting feature representations naturally benefit downstream CTR estimation because the second stage directly inherits the pre-trained scoring function without any architectural modification. The core design is a single set of per-field learnable log-difficulty parameters $\{s^i\}_{i=1}^N$, each a scalar in $\mathbb{R}$, jointly learned with the denoising network $p_\theta$. This unified signal simultaneously drives two coordinated mechanisms that address the difficulty imbalance at the loss and attention levels.

\subsection{Heterogeneous Feature Type Encoding}\label{sec:encoding}

We partition all features into four groups based on their modality:

\textbf{ID fields} ($\mathcal{F}^{ID}$): high-cardinality user and item identity fields. Each value is mapped to a trainable embedding $e^i \in \mathbb{R}^d$.

\textbf{Categorical attribute fields} ($\mathcal{F}^{cat}$): sparse categorical fields with moderate cardinality ($10^1$--$10^4$). Values are mapped to shared category embeddings.

\textbf{Numerical fields} ($\mathcal{F}^{num}$): continuous-valued features. Each value $v^i \in \mathbb{R}$ is discretized into $B$ uniformly spaced bins via $b^i = \min\bigl(\lfloor B \cdot \hat{F}^i(v^i) \rfloor,\, B-1\bigr)$, where $\hat{F}^i$ is the empirical cumulative distribution function of field $i$ computed over the training set.

\textbf{Sequence fields} ($\mathcal{F}^{seq}$): variable-length behavioral histories. Each history is encoded by a lightweight Transformer \cite{transformer} over item embeddings, projected to a fixed-size field embedding $e^i \in \mathbb{R}^d$ via mean pooling and a linear layer.

After type-specific encoding, all $N$ field embeddings $\{e^i\}_{i=1}^N$ are concatenated and passed to the diffusion backbone.

\subsection{Self-Balancing Loss}\label{sec:balance}

Given the field-specific denoising predictions, we decompose the total generation loss into per-field reconstruction terms:
\begin{gather}
\ell^i = -\mathbb{E}_t\!\left[\log p_\theta(f_0^i | \{\mathbf{x}_t^j\}_{j=1}^N, t)\right].
\label{eq:fieldloss}
\end{gather}

\textbf{Multi-Task Likelihood.} Rather than heuristically reweighting per-field losses, we derive the aggregation rule from a principled maximum-likelihood framework. Treating each feature field as an independent reconstruction task, we introduce a per-field homoscedastic uncertainty $(\sigma^i)^2 > 0$ that captures the task-level noise inherent to reconstructing field $i$, independent of any particular input sample. Following the multi-task uncertainty weighting principle \cite{uncertainty_mtl}, we model the likelihood of field $i$ as a Gaussian over the reconstruction residual:
\begin{gather}
p(f_0^i | \{\mathbf{x}_t^j\}, t; \sigma^i) = \mathcal{N}\!\bigl(f_0^i;\, \mu_\theta^i(\{\mathbf{x}_t^j\}, t),\, (\sigma^i)^2\bigr),
\label{eq:gaussian}
\end{gather}
where $\mu_\theta^i$ is the model's predicted sufficient statistic for field $i$. For discrete reconstruction tasks, the same functional form emerges from a scaled softmax likelihood under the standard temperature-scaling approximation \cite{uncertainty_mtl}. Because the $N$ fields are conditionally independent given the shared network output, the joint likelihood factorises as:
\begin{gather}
p\bigl(\{f_0^i\}_{i=1}^N \,|\, \{\mathbf{x}_t^j\}, t; \{\sigma^i\}\bigr) = \prod_{i=1}^N p(f_0^i | \{\mathbf{x}_t^j\}, t; \sigma^i).
\label{eq:joint}
\end{gather}

\textbf{Log-Likelihood Objective.} Maximising the joint likelihood is equivalent to maximising its logarithm. Substituting Eq.~\eqref{eq:gaussian} into Eq.~\eqref{eq:joint} and dropping constants yields the log-likelihood:
\begin{gather}
\log p = -\sum_{i=1}^N \left[ \frac{1}{(\sigma^i)^2} \, \ell^i + \log \sigma^i \right],
\label{eq:loglike}
\end{gather}
where $\ell^i$ is the per-field reconstruction loss defined in Eq.~\eqref{eq:fieldloss}. We therefore minimise the negative log-likelihood with respect to both the network parameters $\theta$ and the task uncertainties $\{\sigma^i\}$:
\begin{gather}
\mathcal{L}_{\text{ML}}(\theta, \{\sigma^i\}) = \sum_{i=1}^N \left[ \frac{1}{(\sigma^i)^2} \, \ell^i + \log \sigma^i \right].
\label{eq:ml}
\end{gather}
The weight $1/(\sigma^i)^2$ down-weights tasks with high uncertainty, while the regulariser $\log \sigma^i$ prevents the model from trivially ignoring any field by pushing $\sigma^i \to \infty$.

\textbf{Gradient Analysis.} Differentiating Eq.~\eqref{eq:ml} with respect to $\theta$ reveals how the uncertainty parameters shape the network gradients:
\begin{gather}
\nabla_\theta \mathcal{L}_{\text{ML}} = \sum_{i=1}^N \frac{1}{(\sigma^i)^2} \, \nabla_\theta \ell^i.
\label{eq:gradtheta}
\end{gather}
Each field contributes to the total gradient in proportion to the inverse of its task uncertainty. Fields with large $(\sigma^i)^2$ (high uncertainty, hard to reconstruct) receive small weights and contribute weakly; fields with small $(\sigma^i)^2$ (low uncertainty, easy to reconstruct) dominate the gradient. This is precisely the imbalance phenomenon described in \S\ref{sec:motivation}. To discover the optimal uncertainties, we differentiate Eq.~\eqref{eq:ml} with respect to $\sigma^i$ and set to zero:
\begin{gather}
\frac{\partial \mathcal{L}_{\text{ML}}}{\partial \sigma^i} = -\frac{2\ell^i}{(\sigma^i)^3} + \frac{1}{\sigma^i} = 0
\quad\Longrightarrow\quad
(\sigma^i)^2 = 2\ell^i.
\label{eq:optsigma}
\end{gather}
At the optimal uncertainty, the equilibrium weight is $1/(\sigma^i)^2 = 1/(2\ell^i)$, inversely proportional to the reconstruction loss.

\textbf{Log-Variance Reparameterisation.} Directly regressing $\sigma^i$ is numerically unstable because Eq.~\eqref{eq:ml} involves division by $(\sigma^i)^3$ in the gradient, which can vanish. Following \cite{uncertainty_mtl}, we reparameterise with the log-variance $s^i := \log (\sigma^i)^2 \in \mathbb{R}$. This yields $(\sigma^i)^2 = \exp(s^i)$ and $\log \sigma^i = s^i/2$, transforming the objective to:
\begin{gather}
\mathcal{L}_{bal} = \sum_{i=1}^N \left[ \exp(-s^i) \cdot \ell^i + \frac{s^i}{2} \right].
\label{eq:bal}
\end{gather}
The mapping $s^i \mapsto \exp(-s^i)$ resolves to the positive domain automatically, ensuring well-formed positive weights without constrained optimisation. The gradient with respect to $s^i$ is:
\begin{gather}
\frac{\partial \mathcal{L}_{bal}}{\partial s^i} = \frac{1}{2} - \exp(-s^i) \, \ell^i,
\label{eq:grads}
\end{gather}
which shares the same equilibrium condition as Eq.~\eqref{eq:optsigma}: $\exp(-s^i) = 1/(2\ell^i)$.

\textbf{Self-Balancing Equilibrium.} For fixed network parameters $\theta$, setting Eq.~\eqref{eq:grads} to zero gives the equilibrium:
\begin{gather}
\exp(-s^i) = \frac{1}{2\ell^i}.
\label{eq:equilibrium}
\end{gather}
The equilibrium loss weight is therefore inversely proportional to the current reconstruction loss. Fields with high $\ell^i$ (hard fields) retain large weights; fields with low $\ell^i$ (easy fields) are down-weighted.

\textbf{Equilibrium Stability and Uniqueness.} We establish that the equilibrium in Eq.~\eqref{eq:equilibrium} is not only necessary but sufficient and stable. The second derivative of $\mathcal{L}_{bal}$ with respect to $s^i$ is:
\begin{gather}
\frac{\partial^2 \mathcal{L}_{bal}}{\partial (s^i)^2} = \exp(-s^i) \, \ell^i > 0 \quad \text{for all } s^i \in \mathbb{R} \text{ and } \ell^i > 0.
\end{gather}
Because the objective is strictly convex in each $s^i$, the equilibrium $(s^i)^* = \log(2\ell^i)$ is the \emph{unique global minimum}. Under gradient descent with learning rate $\eta$, the dynamics around the equilibrium linearise as $\delta_{t+1} = (1 - \eta/2)\,\delta_t$ where $\delta_t = s^i_t - (s^i)^*$. For any standard learning rate $\eta \in (0, 4)$, we have $|1 - \eta/2| < 1$, guaranteeing exponential convergence to the equilibrium. This stability ensures that $s^i$ reliably tracks the moving target $(s^i)^*$ as $\ell^i$ evolves during training.

\textbf{Dynamic Gradient Reallocation.} During training, $\theta$ evolves and $\ell^i$ changes continuously. Consider the update of $s^i$ under gradient descent:
\begin{gather}
s^i_{t+1} = s^i_t - \eta \left( \frac{1}{2} - \exp(-s^i_t)\,\ell^i_t \right).
\end{gather}
If field $i$ becomes harder ($\ell^i$ increases), the parenthetical term becomes negative at the old equilibrium, driving $s^i$ downward and increasing the weight $\exp(-s^i)$. Conversely, if the field becomes easier ($\ell^i$ decreases), the term becomes positive, driving $s^i$ upward and decreasing the weight. This creates a negative feedback loop: the mechanism automatically reallocates gradient budget toward fields that currently need it most. Early in training all fields are hard and $s^i \approx 0$, so all weights are near unity; as easy fields converge, their $s^i$ grow and the mechanism progressively shifts capacity toward the remaining hard fields.

\textbf{Comparison with Gradient-Based Balancing.} Existing methods such as GradNorm \cite{gradnorm}, PCGrad \cite{pcgrad}, and MGDA \cite{mgda} operate at the gradient level, requiring per-task gradient computation, projection, or Pareto optimisation. These incur significant overhead and are designed for discrete task boundaries. The self-balancing loss operates entirely at the loss level: $\{s^i\}$ are scalar parameters updated by standard backpropagation, introducing only $N$ additional scalars. Furthermore, gradient-based methods define difficulty through gradient norms or conflict angles, which are noisy early in training. The self-balancing loss instead accumulates per-field difficulty through $\{s^i\}$ across the entire training trajectory, providing a smoother, more stable signal that adapts automatically as the feature distribution evolves.

\subsection{Difficulty-Guided Denoising}\label{sec:attention}

The same learned $\{s^i\}$ address the attention-level imbalance. Within each HSTU layer of the denoising network, the attention query for field $i$ is modulated by:
\begin{gather}
\tilde{q}^i = \exp\!\left(-\frac{s^i}{2}\right) \cdot q^i
\end{gather}
where $q^i = W_Q e^i$ is the standard linear query projection. The scaling factor $\exp(-s^i/2)$ is the inverse of the standard deviation of the homoscedastic difficulty---it is large for hard fields (small $s^i$) and small for easy fields (large $s^i$). This suppresses the attention influence of converged easy fields and amplifies the cross-field information flow toward hard fields \emph{without introducing any new parameters}: the query weight matrix $W_Q$ is shared with the standard HSTU, and $\{s^i\}$ are already learned for the self-balancing loss.

The modulated attention output for field $i$ is:
\begin{gather}
\text{Attn}^i = \text{softmax}\!\left(\frac{\tilde{q}^i {\mathbf{K}}^\top}{\sqrt{d}}\right)\mathbf{V}
\end{gather}
where $\mathbf{K}, \mathbf{V}$ are the standard key and value matrices. The scaling shifts the attention distribution so that fields with lower difficulty (easy fields) attend more uniformly and contribute less sharply to the pooled representation, while hard fields with higher difficulty attend more selectively to the most relevant context.

\textbf{Why Query Modulation?} The choice to modulate the query $q^i$ rather than the key $k^i$ or value $v^i$ is deliberate. In self-attention, the query of field $i$ determines \emph{how aggressively field $i$ gathers information from all other fields}. Suppressing the query of an easy field reduces its outgoing attention mass, making it a more passive recipient rather than an active information aggregator---precisely the desired behavior, since easy fields have already converged and should not drive the denoising computation. In contrast, modulating the key would control how much other fields attend to field $i$, and modulating the value would scale field $i$'s contribution to other fields' outputs; neither directly captures the goal of controlling field $i$'s own information-gathering behavior.

\textbf{Coefficient Alignment with Self-Balancing Loss.} The choice of $\exp(-s^i/2)$ rather than $\exp(-s^i)$ is deliberate. In the self-balancing loss, the loss weight $\exp(-s^i)$ acts as a multiplier on the scalar loss $\ell^i$. In the attention mechanism, scaling the query by $\exp(-s^i/2)$ directly modulates the magnitude of the pre-softmax logits: hard fields with small $s^i$ receive larger query norms, producing sharper attention distributions that focus aggressively on the most relevant context keys; easy fields with large $s^i$ receive smaller query norms, yielding softer, more uniform attention that reduces their influence on the pooled representation. Consequently, the attention modulation and the loss reweighting are qualitatively aligned---both suppress easy fields and amplify hard fields---so the two components reinforce rather than conflict with each other.

\subsection{Generative Training Objective}

\subsubsection{Pre-Training Objective}
HeteGenCTR follows the same coupled two-stage training paradigm as DGenCTR. The pre-training objective reconstructs all fields including the click label, which is treated as an additional feature field within the sample. HeteGenCTR improves upon DGenCTR by replacing the uniform loss aggregation over the $N$ input feature fields with the self-balancing objective; the label field continues to be reconstructed during pre-training exactly as in DGenCTR, ensuring the scoring function remains aligned with the downstream CTR objective. The total pre-training loss is therefore:
\begin{gather}
\mathcal{L}_{\text{pretrain}} =  \sum_{i=1}^{N} \left[\exp(-s^i) \cdot \ell^i + \frac{s^i}{2}\right]
\end{gather}
Both $\theta$ and $\{s^i\}$ are updated by gradient descent on $\mathcal{L}_{\text{pretrain}}$. 

\begin{algorithm}[t]
\small
\caption{HeteGenCTR Generative Pre-Training}
\label{alg:hetegenctr}
\KwIn{Training dataset $\mathcal{D}$; diffusion steps $T$; discretisation bins $B$}
\KwOut{Trained denoising network $p_\theta$; log-difficulty parameters $\{s^i\}$}

Initialize $p_\theta$ and $\{s^i = 0\}_{i=1}^N$\;
\tcp{--- Generative Pre-Training ---}
\For{each training batch $\{\mathbf{F}_n\}$}{
  Encode each field $f^i_n$ via its type-specific encoder to obtain $e^i_n$\;
  Sample timestep $t \sim \text{Uniform}(1, T)$\;
  For each field $i$: sample $\mathbf{x}_t^{i,n} \sim q(\cdot | f^i_n)$ using per-field schedule $\gamma^i(t)$\;
  Compute $\tilde{q}^{i,n} = \exp(-s^i/2) \cdot W_Q e^i_n$ for difficulty-guided attention\;
  Predict $p_\theta(f^i_0 | \{\mathbf{x}_t^{j,n}\}_j, t)$ for all fields $i$ using modulated attention\;
  Compute per-field reconstruction loss $\ell^i$ for all input feature fields\;
  $\mathcal{L} \leftarrow \sum_i \left[\exp(-s^i) \cdot \ell^i + s^i / 2\right]$ \tcp*{input-feature losses only; label prediction preserved as in DGenCTR}
  Update $\theta$ and $\{s^i\}$ via gradient descent on $\mathcal{L}$\;
}
\end{algorithm}

\subsubsection{CTR-Targeted Fine-Tuning}
In the second stage, the pre-trained scoring function and all network parameters are directly inherited and fine-tuned for precise CTR prediction as DGenCTR \cite{dgenctr}. Importantly, the inherited scoring function and its training objective remain unchanged; the only difference is that the pre-trained network parameters now encode higher-quality, heterogeneity-aware feature representations, as the self-balancing mechanism ensures that hard fields receive sufficient gradient signal throughout stage-one training. Because the label-aware generative objective in stage one is mathematically equivalent to a CTR calibration loss (Eq.~\eqref{eq:label}), the scoring function learned during pre-training is already aligned with the downstream CTR objective. The second-stage fine-tuning therefore minimizes the standard binary cross-entropy:
\begin{gather}
\mathcal{L}_{\text{SFT}} = -y \log \sigma(z) - (1-y) \log \bigl(1 - \sigma(z)\bigr) \\
 \quad z = \mathcal{F}(y=1|\mathbf{F}) - \mathcal{F}(y=0|\mathbf{F})
\label{eq:sft}
\end{gather}

\section{Experiments}

We evaluate HeteGenCTR on large-scale benchmark datasets to address the following research questions:\\
\textbf{$\bullet$ RQ1:} Does HeteGenCTR consistently improve downstream CTR prediction performance compared to state-of-the-art generative and discriminative baselines?\\
\textbf{$\bullet$ RQ2:} What is the contribution of each component of HeteGenCTR to overall performance?\\
\textbf{$\bullet$ RQ3:} How do the $\{s^i\}$ evolve during training, and what per-field loss weights and attention scaling factors do they converge to?\\
\textbf{$\bullet$ RQ4:} Does the self-balancing mechanism effectively improve per-field generation quality and downstream CTR performance?\\
\textbf{$\bullet$ RQ5:} How sensitive is HeteGenCTR to pre-training hyperparameters (pre-training epochs $N_{\text{pretrain}}$ and diffusion steps $T$), and what is its practical training overhead?\\
\textbf{$\bullet$ RQ6:} Does HeteGenCTR provide disproportionate gains for cold-start and sparse users?

\subsection{Datasets}

We evaluate on four large-scale public benchmark datasets and one industrial dataset. Statistics are summarized in Table~\ref{datasets}.

\textbf{$\bullet$ Criteo}\footnote{http://labs.criteo.com/downloads/download-terabyte-click-logs/}. A canonical public benchmark for display advertising CTR prediction. It contains 13 dense numerical features and 26 high-cardinality categorical features, providing a balanced test of methods on mixed feature types \cite{criteo}.

\textbf{$\bullet$ Avazu}\footnote{http://www.kaggle.com/c/avazu-ctr-prediction}. A widely used mobile advertising CTR benchmark consisting of 10 consecutive days of ad click logs with 23 categorical feature fields. Its features are overwhelmingly sparse, making it a strong test of robustness under high sparsity \cite{avazu}.

\textbf{$\bullet$ KDD12}\footnote{http://www.kddcup2012.org/c/kddcup2012-track2/data}. A search advertising dataset derived from user session logs, containing 11 categorical fields describing user--query--ad interactions. It is characterized by extreme class imbalance (CTR below 5\%), challenging models to extract signal from positive events.

\textbf{$\bullet$ Amazon Product Reviews (Electronics)}. A product recommendation dataset \cite{amazon} that combines user ID, item ID, product category, brand, and historical behavioral sequence fields. The inclusion of sequential features alongside IDs and categories makes it the most heterogeneous public benchmark in our suite, directly testing HeteGenCTR's ability to handle diverse field types.

\textbf{$\bullet$ Industrial Dataset}. A proprietary dataset collected from a large-scale e-commerce platform's online display advertising system. It comprises 68 feature fields spanning numerical attributes, low- and high-cardinality categorical fields, user/item IDs, and multi-length behavioral sequences. The training set contains impressions from the last 20 days; the test set comprises held-out exposure samples from the subsequent day. Its high feature type diversity and realistic long-tail distribution make it the most challenging evaluation setting.

\begin{table}[t]
  \small
  \centering
  \begin{tabular}{ccccc}
    \hline
    Dataset & \# Fields & \# Impressions & \# Positive & Types \\
    \hline
    Criteo & 39 & 45M & 26\% & num/cat \\
    Avazu & 23 & 40M & 17\% & cat \\
    KDD12 & 11 & 60M & 4.5\% & cat \\
    Amazon & 18 & 12M & 8.3\% & ID/cat/seq \\
    Industrial & 68 & 513M & 2.5\% & all \\
    \hline
  \end{tabular}
  \caption{Statistics of datasets. ``Types'' indicates the feature field types present in each dataset.}
  \label{datasets}
\end{table}

\subsection{Competitors}

\begin{table*}[t]
  \caption{Prediction performance of CTR models on five datasets. * indicates $p < 0.05$ in significance test.}
  \begin{tabular}{c|cc|cc|cc|cc|cc}
    \toprule
    \multirow{2}{*}{\diagbox{Method}{Dataset}} &
      \multicolumn{2}{c|}{Criteo} &
      \multicolumn{2}{c|}{Avazu} &
      \multicolumn{2}{c|}{KDD12} &
      \multicolumn{2}{c|}{Amazon} &
      \multicolumn{2}{c}{Industrial} \\
    \cmidrule(lr){2-11}
    & AUC & Logloss & AUC & Logloss & AUC & Logloss & AUC & Logloss & AUC & Logloss \\
    \midrule
    DeepFM & 0.7692 & 0.4713 & 0.7756 & 0.4469 & 0.7933 & 0.1422 & 0.8021 & 0.2314 & 0.7785 & 0.0852 \\
    DCN & 0.7703 & 0.4703 & 0.7762 & 0.4458 & 0.7941 & 0.1426 & 0.8035 & 0.2301 & 0.7792 & 0.0851 \\
    AutoInt & 0.7695 & 0.4710 & 0.7748 & 0.4473 & 0.7928 & 0.1429 & 0.8019 & 0.2318 & 0.7823 & 0.0847 \\
    FiBiNet & 0.7732 & 0.4691 & 0.7759 & 0.4456 & 0.7968 & 0.1402 & 0.8043 & 0.2287 & 0.7825 & 0.0844 \\
    MaskNet & 0.7882 & 0.4644 & 0.7813 & 0.4415 & 0.8012 & 0.1381 & 0.8074 & 0.2265 & 0.7846 & 0.0831 \\
    PEPNet & 0.7981 & 0.4498 & 0.7944 & 0.4402 & 0.8041 & 0.1370 & 0.8096 & 0.2247 & 0.7904 & 0.0817 \\
    HSTU & 0.7993 & 0.4483 & 0.7902 & 0.4403 & 0.8087 & 0.1358 & 0.8112 & 0.2235 & 0.7926 & 0.0814 \\
    \midrule
    GenCTR & 0.8003 & 0.4472 & 0.7931 & 0.4391 & 0.8091 & 0.1354 & 0.8118 & 0.2231 & 0.7934 & 0.0810 \\
    DGenCTR & 0.8024 & 0.4459 & 0.7947 & 0.4383 & 0.8106 & 0.1348 & 0.8127 & 0.2219 & 0.7948 & 0.0806 \\
    SGCTR & 0.8031 & 0.4455 & 0.7953 & 0.4378 & 0.8118 & 0.1342 & 0.8139 & 0.2211 & 0.7956 & 0.0804 \\
    \midrule
    \textbf{HeteGenCTR} & \textbf{0.8048*} & \textbf{0.4445*} & \textbf{0.7962*} & \textbf{0.4373*} & \textbf{0.8127*} & \textbf{0.1334*} & \textbf{0.8157*} & \textbf{0.2188*} & \textbf{0.7974*} & \textbf{0.0799*} \\
    \bottomrule
  \end{tabular}
  \label{results}
  \vspace{-0.1cm}
\end{table*}

\textbf{1) Discriminative Models}: DeepFM \cite{deepfm}, DCN \cite{dcn}, AutoInt \cite{autoint}, FiBiNet \cite{fibinet}, MaskNet \cite{masknet}, PEPNet \cite{pepnet}, and HSTU \cite{hstu}. \textbf{2) Generative Models}: GenCTR \cite{genctr}, DGenCTR \cite{dgenctr}, and SGCTR \cite{sgctr}.

\textbf{Implementation Details}. All models are implemented in TensorFlow \cite{tensorflow} and trained on 8 NVIDIA A100 GPUs using the Adam optimizer \cite{adam} with Xavier initialization \cite{xavier}. The default activation function is ReLU. Embedding dimension is 32 for public datasets and 8 for Industrial. Batch size is 4096. Learning rates are searched in $\{3\text{e-}3, \ldots, 1\text{e-}5\}$ and $L_2$ regularization in $\{3\text{e-}6, \ldots, 0\}$. For HeteGenCTR, the diffusion process uses $T=100$ timesteps with per-field cosine noise schedules inherited from DGenCTR. The generative pre-training stage runs for $N_{\text{pretrain}}=10$ epochs by default. Numerical discretization uses $B=100$ bins. All log-difficulty parameters $\{s^i\}$ are initialized to 0. The denoising network follows the HSTU architecture following DGenCTR, as its specific design is orthogonal to our core contribution. While HeteGenCTR inherits DGenCTR's discrete diffusion framework for generative pre-training, its inference pipeline follows SGCTR's masked generative paradigm for efficient CTR prediction at serving time.

\textbf{Evaluation Metrics}. We adopt AUC (Area Under ROC Curve) and LogLoss (binary cross-entropy) as primary evaluation metrics, following standard CTR prediction practice.

\subsection{Comparison with Baselines (RQ1)}

Table \ref{results} presents the overall prediction performance across all five datasets. HeteGenCTR consistently outperforms all baselines on every dataset, achieving statistically significant improvements ($p < 0.05$) over the strongest generative baseline SGCTR.

Several observations are noteworthy. First, generative CTR baselines (GenCTR, DGenCTR, SGCTR) consistently outperform the best discriminative models, confirming the value of generative pre-training: reconstruction provides dense supervision across all feature fields, whereas discriminative training receives only a single binary signal per sample. Second, HeteGenCTR achieves a further consistent improvement over all generative baselines. Existing generative methods apply uniform loss weights, causing easy fields to dominate gradients and leaving high-cardinality ID and sequence fields underfit. HeteGenCTR's self-balancing loss reallocates gradient budget toward hard fields, while difficulty-guided attention prevents converged easy fields from monopolizing cross-field information flow; the combination produces higher-quality reconstructions for the fields that carry the strongest personalization signal. Third, the gains are most pronounced on Amazon and Industrial---the datasets with the greatest feature type diversity. This is expected from the algorithmic design: when heterogeneity is high, the easy-hard gap widens and the uniform-weight baseline becomes more suboptimal, giving the self-balancing mechanism a larger corrective effect. On Criteo and Avazu, the difficulty gap is inherently smaller, so the rebalancing gain is reduced but remains consistent and statistically significant. Fourth, LogLoss improvements corroborate the AUC gains across all datasets: better reconstruction of hard fields yields more informative embeddings that improve not only ranking quality but also probability calibration.

\subsection{Ablation Study (RQ2)}
\begin{itemize}
\item \textbf{HeteGenCTR-FIX}: disables the self-balancing loss, while retaining difficulty-guided attention modulation. The log-difficulty parameters $\{s^i\}$ are still learned, but their gradients originate solely from the attention-modulation pathway.
\item \textbf{HeteGenCTR-STD}: disables difficulty-guided attention modulation, while retaining the self-balancing loss. The $\{s^i\}$ are learned from the loss-reweighting gradients only.
\end{itemize}

\begin{figure}[t]
  \centering
  \includegraphics[width=\linewidth]{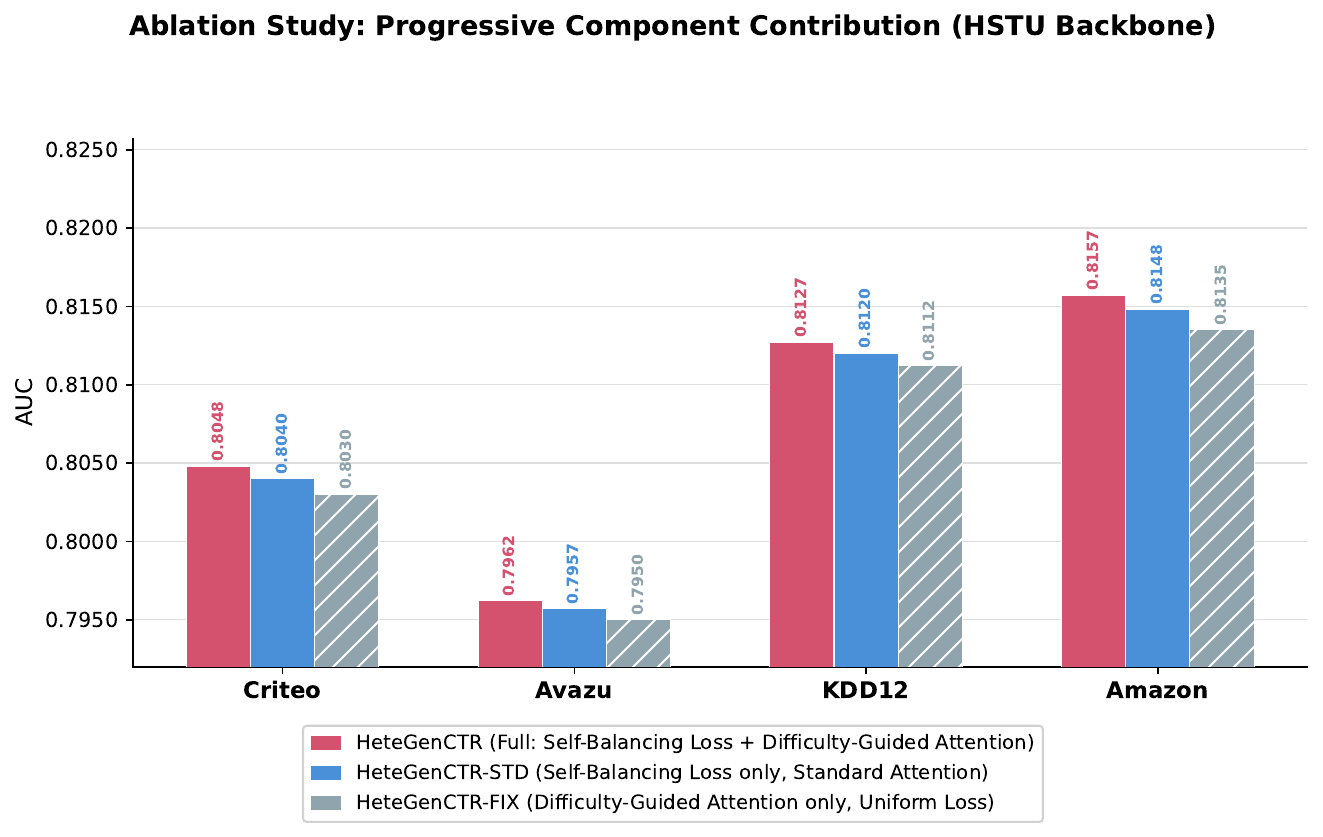}
  \caption{Ablation study results of two variants. }
  \label{abla}
\end{figure}

Figure \ref{abla} presents the ablation results. FIX retains difficulty-guided attention modulation but replaces the self-balancing loss with uniform field weights. Without adaptive gradient reallocation, easy fields still dominate the training budget, and the attention modulation alone cannot fully compensate for the poor embeddings learned under uniform supervision. Restoring the self-balancing loss (Full vs.\ FIX) yields substantial AUC gains on every dataset, confirming that loss-level reweighting is the primary driver of HeteGenCTR's improvement. The gain is largest on Amazon and KDD12, where feature heterogeneity is highest. STD uses the self-balancing loss but reverts to standard unmodulated attention. Even with balanced gradients, the HSTU attention can still be disproportionately influenced by easy fields at the representation level. Adding difficulty-guided attention modulation (Full vs.\ STD) provides a further consistent improvement: suppressing easy-field attention queries with $\exp(-s^i/2)$ allows hard fields to exert stronger cross-field influence during denoising, producing more coherent feature reconstructions for high-cardinality fields. The coefficient $\exp(-s^i/2)$ is derived to align the attention-level weighting with the loss-level weighting, ensuring both mechanisms reinforce the same difficulty signal. The progressive improvement pattern FIX $<$ STD $<$ Full across all settings confirms that each mechanism provides independent, additive value, and the unified $\{s^i\}$ signal successfully coordinates both components.

\subsection{Difficulty Parameter Analysis (RQ3)}\label{sec:rq3}

\textbf{Difficulty evolution.} Figure \ref{hyper}(a) visualizes the evolution of $\exp(s^i)$ (the difficulty magnitude, proportional to $1/\text{weight}$) for representative fields of each type during training on the Industrial dataset. Early in training, all $\exp(s^i) \approx 1$ (corresponding to $s^i=0$ initialization). As training progresses, $\exp(s^i)$ for numerical and low-cardinality categorical fields increases sharply, indicating that the model has mastered these fields and their effective loss weight has been reduced. In contrast, $\exp(s^i)$ for ID and sequence fields increases much more slowly, reflecting persistently high reconstruction difficulty and maintaining strong gradient signals for these fields. This differential evolution confirms that the self-balancing mechanism behaves as intended.

\textbf{Converged difficulty weights.} Figure \ref{hyper}(b) shows the converged per-field effective loss weights $\exp(-s^i)$ and attention scaling factors $\exp(-s^i/2)$ for the Industrial dataset. ID and sequence fields converge to significantly smaller $\exp(-s^i)$ (i.e., larger loss weight $1/\exp(-s^i)$), while numerical and low-cardinality categorical fields receive smaller effective weights. Crucially, the attention scaling $\exp(-s^i/2)$ follows the same rank ordering as the loss weights, confirming that both mechanisms are driven by the same learned signal and remain mutually consistent throughout training.

\subsection{Generation Quality and CTR (RQ4)}

\textbf{Per-field generation quality.} We measure reconstruction accuracy (categorical accuracy for discrete fields, binned accuracy for numerical fields) of HeteGenCTR vs.\ DGenCTR on a held-out validation set. HeteGenCTR achieves $+4.3\%$ absolute improvement for ID fields and $+2.8\%$ for sequence fields, while improvements for categorical and numerical fields are smaller ($+1.1\%$ and $+0.9\%$ respectively). This confirms that the self-balancing mechanism specifically improves generation quality for the field types that are hardest to reconstruct and most informative for downstream CTR.

\textbf{Downstream impact by field type.} We construct controlled variants that apply HeteGenCTR's self-balancing mechanism for one field type at a time while using DGenCTR's uniform generation for all others. Results on the Industrial dataset confirm that ID and sequence fields are the primary contributors to the overall AUC gain. Applying self-balancing to ID fields alone yields the largest single-type contribution, followed by sequence fields, with categorical and numerical fields providing smaller but consistent gains. The per-type effects are sub-additive when combined, as expected given the shared representation space, but the full HeteGenCTR achieves the maximum overall gain, validating that heterogeneous self-balancing improves downstream CTR by specifically enhancing generation quality for the most difficult and informative field types.

\subsection{Pre-training Sensitivity and Cost (RQ5)}

\begin{figure*}[t]
  \centering
  \includegraphics[width=\linewidth]{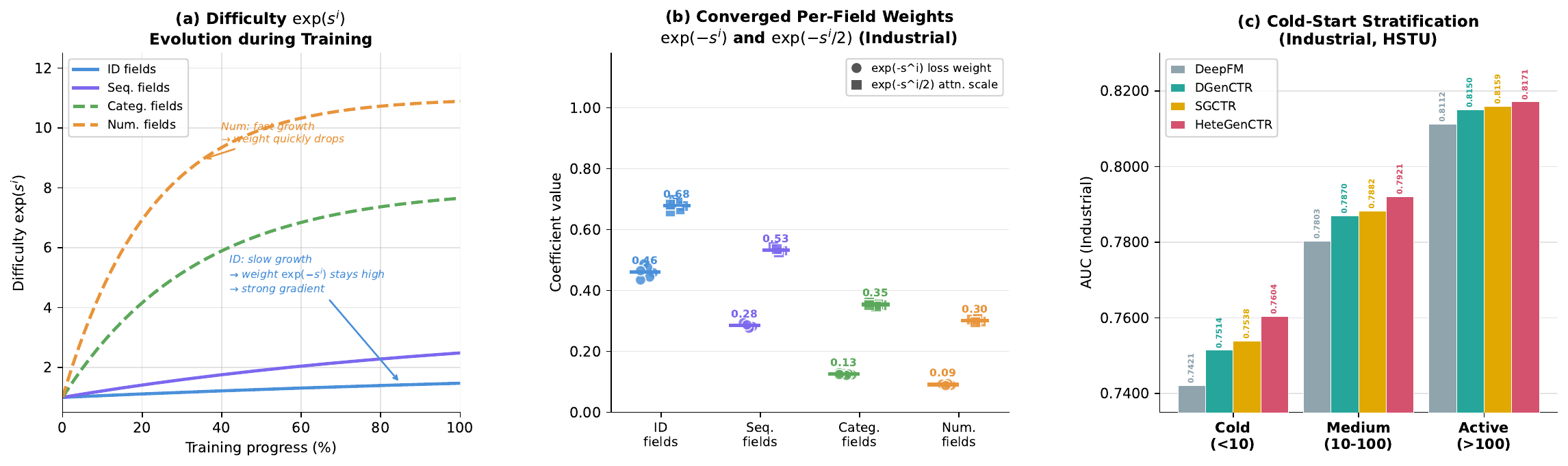}
  \caption{Analysis of learned difficulty parameters. (a) Evolution of $\exp(s^i)$ for representative fields. (b) Converged per-field loss weights $\exp(-s^i)$ and attention scaling factors $\exp(-s^i/2)$ across field types. (c) AUC improvement by user activity stratum.}
  \label{hyper}
  \vspace{-0.4cm}
\end{figure*}

Following the two-stage generative CTR paradigm, HeteGenCTR's downstream performance depends on the quality of generative pre-training in stage one. We analyze sensitivity to two key pre-training hyperparameters on the Industrial dataset.

\begin{figure}[t]
  \centering
  \includegraphics[width=\linewidth]{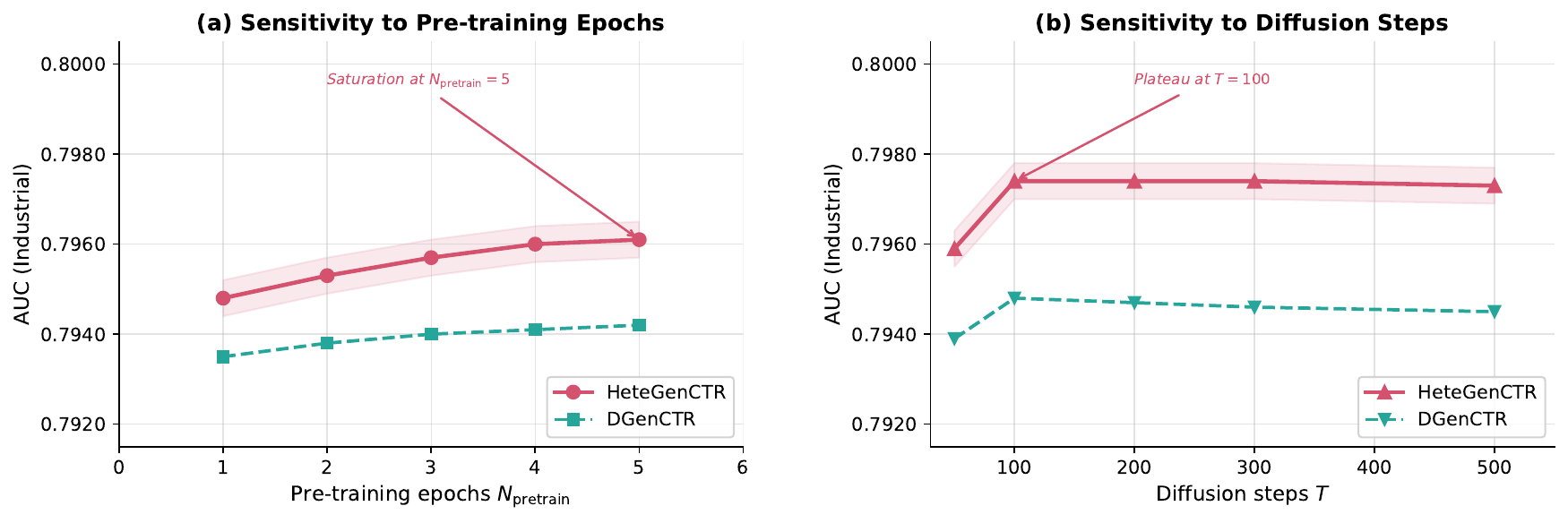}
  \caption{Pre-training sensitivity analysis.}
  \label{sens}
  \vspace{-0.2cm}
\end{figure}

\textbf{Sensitivity to pre-training epochs $N_{\text{pretrain}}$.} We vary the number of generative pre-training epochs before CTR-targeted fine-tuning. Figure~\ref{sens}(a) reports downstream AUC. Performance improves monotonically with more pre-training epochs, rising from $0.7948$ at 1 epoch to $0.7961$ at 5 epochs, with the largest marginal gain occurring between 1 and 3 epochs ($+0.0009$). This confirms that HeteGenCTR's self-balancing mechanism requires sufficient pre-training iterations to discover and stabilize the per-field difficulty estimates; with too few epochs, the $\{s^i\}$ have not converged and the gradient reallocation remains suboptimal. The diminishing returns beyond 3 epochs indicate that the difficulty parameters have largely reached a stable equilibrium on this dataset.

\textbf{Sensitivity to diffusion steps $T$.} We vary the number of discrete diffusion timesteps in the forward and reverse processes. Figure~\ref{sens}(b) reports downstream AUC. Performance improves from $T=50$ to $T=100$ ($0.7959 \rightarrow 0.7974$), then plateaus from $T=200$ through $T=500$ ($0.7974$, $0.7974$, and $0.7973$). This pattern aligns with DGenCTR's observation that a moderate number of diffusion steps provides sufficient noise granularity for effective generative pre-training, while excessive steps introduce unnecessary computational cost without downstream benefit.

\textbf{Training overhead analysis.} Table~\ref{overhead} reports wall-clock training time for the two-stage pipeline normalized to a single-stage DeepFM baseline on the Industrial dataset.

HeteGenCTR's pre-training cost is comparable to DGenCTR. The additional overhead originates from two lightweight additions: (1) the per-field log-difficulty parameter updates,; and (2) the difficulty-guided attention computation, which replaces the standard query projection with a scaled variant requiring no extra parameters. Both additions are negligible compared to the HSTU backbone. The second-stage fine-tuning cost is identical across all generative methods ($0.4\times$), and the generative pre-training phase is offline, so HeteGenCTR incurs no additional online serving latency.

\subsection{Cold-Start and Sparse User Analysis (RQ6)}

A core motivation of HeteGenCTR is that feature heterogeneity causes the greatest harm for instances where high-cardinality fields are sparsely observed---cold-start users and long-tail items. We verify this by stratifying the Industrial test set by user activity level.

\textbf{User activity stratification.} We partition users into three strata based on historical click count: cold users ($<$10 clicks), medium-active users (10--100 clicks), and active users ($>$100 clicks). Table~\ref{coldstart} reports AUC within each stratum.HeteGenCTR's improvement over SGCTR monotonically decreases with user activity: $+0.0066$ for cold users, $+0.0039$ for medium-active, and $+0.0012$ for active users. This confirms the core mechanism: cold users have sparse ID embeddings that are hardest to reconstruct under uniform generation, and the self-balancing mechanism specifically reallocates training capacity toward these high-difficulty ID fields, producing higher-quality feature reconstructions exactly where they are most needed. Figure~\ref{hyper}(c) visualizes this stratum-level improvement, further confirming HeteGenCTR's disproportionate benefit to the most challenging user segments.

\begin{table}[t]
\small
\centering
\begin{tabular}{l|ccc}
\hline
Method & Pre-training & Fine-tuning & Total \\
\hline
DGenCTR    & $1.8\times$ & $0.4\times$ & $2.2\times$ \\
SGCTR      & $2.1\times$ & $0.5\times$ & $2.6\times$ \\
HeteGenCTR & $2.0\times$ & $0.4\times$ & $2.4\times$ \\
\hline
\end{tabular}
\caption{Training overhead relative to DeepFM.}
\label{overhead}
\vspace{-0.1cm}
\end{table}

\begin{table}[t]
\small
\centering
\begin{tabular}{l|ccc|c}
\hline
Method & Cold ($<$10) & Medium & Active ($>$100) & Overall \\
\hline
DeepFM     & 0.7421 & 0.7803 & 0.8112 & 0.7785 \\
DGenCTR    & 0.7514 & 0.7870 & 0.8150 & 0.7948 \\
SGCTR      & 0.7538 & 0.7882 & 0.8159 & 0.7956 \\
\textbf{HeteGenCTR} & \textbf{0.7604} & \textbf{0.7921} & \textbf{0.8171} & \textbf{0.7974} \\
\hline
$\Delta$ vs SGCTR & $+0.0066$ & $+0.0039$ & $+0.0012$ & $+0.0018$ \\
\hline
\end{tabular}
\caption{AUC by user activity stratum on Industrial dataset. }
\label{coldstart}
\end{table}

\textbf{Item long-tail analysis.} Stratifying items by training exposure count, HeteGenCTR achieves AUC improvements of $+0.0091$, $+0.0051$, and $+0.0019$ over SGCTR for tail ($<$100 exposures), torso, and head items respectively, mirroring the cold-user result.

\subsection{Online A/B Testing Results}

To validate real-world efficacy, we deployed HeteGenCTR in a seven-day online A/B test on a large-scale e-commerce platform from May 7 to 13, 2026. The production baseline employs a PEPNet-like \cite{pepnet} discriminative architecture trained without generative pre-training. HeteGenCTR achieved a 4.7\% relative improvement in click-through rate ($p < 0.01$, two-sided $z$-test with randomized user-level traffic splitting), consistent across all seven days.

\textbf{Serving latency.} HeteGenCTR's generative pre-training is entirely an offline stage: no generative component is invoked at serving time. The deployed CTR model is architecturally identical to the baseline, and the 99th-percentile serving latency is within 0.5ms of the baseline, well within the production SLA.

\textbf{Cold-start gain online.} Breaking down by user activity, the experiment shows a 9.2\% CTR improvement for cold-start users (fewer than 10 historical exposures) versus 3.1\% for active users, consistent with the offline stratification analysis.

\section{Conclusions}
We identified and formalized the \emph{generative difficulty imbalance} in generative CTR modeling: the uniform treatment of feature fields in the generation objective causes easy fields to dominate training gradients and suppresses learning for informative but hard-to-reconstruct fields. To resolve this, we proposed HeteGenCTR, which introduces a single set of per-field learnable log-difficulty parameters that simultaneously drive two coordinated mechanisms---self-balancing loss aggregation with a provably stable equilibrium, and difficulty-guided attention modulation with scaling coefficients principled to align with the loss-level weighting. The unified signal ensures that both mechanisms remain consistent with the same learned difficulty signal throughout training, creating a holistic solution to the multi-level imbalance. Experiments confirm consistent significant improvements over state-of-the-art baselines.

\end{document}